\documentclass[a4paper]{article}
\usepackage{graphicx}
\begin{document}

${\bf Solar Neutron Decay Protons observed in November 7, 2004}$
\\
\\
Yasushi Muraki$^{ 1, 5,*}$\footnote{*)correspondence:{\,}muraki@isee.nagoya-u.ac.jp},
 Jose F. Vald$\acute{e}s$-Galicia$^2$, Ernesto Ortiz$^3$, 
Yutaka Matsubara$^1$,
Shoichi Shibata$^4$, Takashi Sako$^{1, 6}$, Satoshi  Masuda$^1$, 
Munetoshi  Tokumaru$^1$, 
Tatsumi Koi$^4$, Akitoshi Ooshima$^{4 ,5}$, Takasuke Sakai$^7$, 
Tsuguya Naito$^8$, and 
Pedro Miranda${^9}$
\footnote{Full author information is available at the end of the article.}\,
\begin{abstract}

We have found an interesting event registered by the solar neutron telescopes 
installed at high mountains in Bolivia (5250 m a.s.l.) and Mexico (4600 m a.s.l.). 
The event was observed November 7th of 2004 in association with a large solar flare 
of magnitude X2.0.  Some features in our registers and in two satellites (GOES 11 and SOHO) 
reveal the presence of electrons and protons as possible products of neutron decay. 
Solar neutron decay protons (sndp) were recorded on board ISEE3 satellite in June 3rd, 1982 . 
On October 19th, 1989, the ground level detectors installed in Goose Bay and Deep River 
revealed the registration of solar neutron decay protons (sndp). 
Therefore this is the second example that such an evidence is registered on the Earth$'$s surface.
\end{abstract}

${\bf key words}$: Solar neutron decay protons, Solar flare, Solar Energetic Particles, Particle acceleration

\section{Introduction}

  Gamma rays and neutrons propagate freely in the interplanetary medium 
when emitted as secondary products of solar explosion events. Therefore they may provide
information regarding the condition of production site and mechanism acceleration. 
Solar neutrons have been observed in space crafts and ground based detectors
 (see, e.g.  Galicia et al. 2009, Dorman 2010, Kamiya et al 2019, Muraki et al. 2020
and reference there in).  In order to push forward this study, we have international solar neutron telescope (SNT) network 
at high mountains in the world.   Evenson et al. (1983) reported the discovery of interplanetary protons 
by the decay of solar flare neutrons.  Shea et al.(1991) found signals in neutron monitors for the event 
on 19 October 1989 that may be interpreted as the detection of relativistic protons that were the decay products of solar neutrons. 
Observing protons produced by the neutron decay in flight, very accurate energy spectra 
may be obtained. These spectra will be close to the source spectra. 

In this paper we report registers obtained at two Solar Neutron Telescopes installed 
at high mountains, combined with the observations of two spacecrafts (GOES 11 and SOHO). 
They could be interested as products of the solar neutrons decay that were produced 
at the 7 November 2004 X2 solar flare. 

The plan of the paper is as follows: in section 2 we give a brief description of the Solar Neutron Telescopes (SNT) 
at Bolivia and Mexico; section 3 is dedicated to a description of the observations in spacecrafts and on the Earth$'$s surface. 
 In section 4, we give a plausible interpretation of the results. Then section 5 is dedicated to convert the fluxes observed 
on earth$'$s surface to fluxes on the top of the atmosphere to compare with the spacecraft observations. 
The sites where solar neutrons may have decayed are analyzed in section 6, to finish with our conclusions in section 7. 

\section{The two mountain detectors located on American continent}

Two solar neutron detectors are located at Mt. Sierra Negra
 (4,600 m, 19.0$^{\circ}$N 97.3$^{\circ}$W) and Mt. Chacaltaya (5,250 m, 16.3$^{\circ}$S 68.1$^{\circ}$W).  
The detector installed at Mt. Chacaltaya in Bolivia is comprised 
of 4 $m^2$ plastic scintillator with 40 cm thickness 
and the detector installed at Mt. Sierra Negra, Mexico is composed 
of the 4 $m^2$ plastic scintillator with 30 cm thickness. 
 In addition, four layers of proportional counters are installed underneath the scintillator for the identification 
of the arrival direction of charged particles (Valdes-Galicia et al., 2004).  
These SNTs were constructed for the detection of solar neutrons.  
Both instruments distinguish neutral incidents from charged particles.

\section{ Main features of the observed events}

1. $\it Signals{\,} recorded{\,} in {\,}the{\,} ground{\,} level{\,} detectors$: increases of the counting rate were recognized 
in the 5-minute-value of the solar neutron telescopes (SNT) located in Bolivia from 15:48 to 16:06 UT, 
and in Mexico from 15:51 to 17:09 UT of November 7th, 2004.  
The statistical significance of each excesses was 3.7$\sigma$ and 12$\sigma$ respectively.  
Both data are shown in Figures 1 and 2. The excesses of signal of the SNTs were observed almost at the same time, 
assuming them to be due to the solar flare of X2.0.  Therefore, the increase cannot be explained by very plausibly 
originated at fluctuations of the background signal. 

2. $\it Signals{\,} recorded{\,} in{\,} the{\,} detectors{\,} onboard {\,}the{\,} satellites$: Figure 3 shows the counting rate of the protons 
with $>$10 MeV observed by the GOES 11 satellite.  An increase is apparently recognized between 15:50 UT and 16:00 UT (16$\sigma$).   
An  appropriate explanation on this bump is that the detector onboard the GOES 11 satellite received solar neutron decay protons (sndp).  
Figure 4 presents another candidate of neutron decay products detected by the electron detector COSTEP 
onboard the SOHO satellite that stays near the Lagrange point L1 (SOHO-COSTEP-EPHIN website). 
The excess was recorded by the COSTEP instrument during 15:58 UT - 16:17 UT.  Solar neutron decay protons (sndp) 
were also observed by the COSTEP.  In the proton channels 4.3-7.8 MeV and 7.8-25 MeV, the excess of the sndp was clearly observed.  
However, in the highest channel 25-35 MeV, the excess was not definitely recognized.  
On the other hand, in the electron channels 0.25-0.70 MeV, 0.67-3.0 MeV and 2.64-10.4 MeV, the excess was clearly see.  
These may be another evidence of neutron decay products via n$\rightarrow p + e^- + \nu_{e}$ process.  
Therefore neutron decay protons and electrons were observed not only inside but also outside of the magnetosphere.  
\section{ Interpretations on the event}

 We make the following hypotheses in order to explain the observations.

(a) The increase of the 5-minute counting rate registered during 15:48-16:07 UT at Chacaltaya was induced by solar neutrons.  
These neutrons were produced by the impulsive flare with X2.0.  One of the most probable production time 
was presumed when the derivative of the GOES X-ray intensity showed the first maximum at 15:47:00 UT (Figure 5).  
Figure 5 shows that the flare reached X2.0 intensity via three step increase. 
The tendency is quite similar to a previous event observed in April 15, 2001 (Muraki et al. 2008).

(b) The increase of the counting rate observed between 15:51 and 17:09 UT at Mt. Sierra Negra was produced
 by solar neutron decay protons (sndp) in the space between the CME front and the Earth. 
 Here we specify the preceding CME as CME1.  Since two large CMEs were produced: one in November 6th at 00:34 UT 
and another in 7th at 16 UT.  Possible acceptable area for the sndp by each detector is pictorially shown in Figure 6.  
The distance D1 in Figure 6 is estimated to be $10^7$ km.  
 First, the sndp produced near the magnetosphere arrived, then the sndp produced near the CME1 transported to the Earth. 
This is due to the fact that charged particles produced in the interplanetary space are transported toward the Earth 
along the inter-planetary magnetic field.  That is one of the reasons why the excess of Mt. Sierra Negra 
continued for 78 minutes and was slightly delayed with respect to the neutron arrival time.  
In comparison with the June 3rd, 1982 event, the sndp of November 7th, 2004 event was observed rather short period ($\sim$1/10). 
 One of the reasons may due to the difference of decay space of the sndp by each detector. An explanation will be given in the next section.

The hypotheses (a) and (b) were introduced for two main reasons.  One arises from the difference of the atmospheric thickness 
of the neutron path length between the two observatories.  At the flare time, there were 200 g/$cm^2$ difference of atmospheric thickness 
between the two observatories for the passage of neutrons (Dorman et al., 1999, Tsuchiya et al. 1999, 2001).  
Therefore the flux at Mt. Sierra Negra should be 7 times less than that of Chacaltaya ($\sim$1/7), 
although both detectors are located at nearly the same altitude. (Let us remind that the Sun was situated 
above South America.)  However, both excesses were detected with nearly the same intensity.

The second reason is based on the difference of the event duration. The signal of Chacaltaya was observed 
for 18 minutes, while the excess of Sierra Negra continued for 78 minutes (as shown by the horizontal arrow in Figure 2).  
Neutrons with 50 MeV energy are expected to arrive to the Earth 18 minutes later than the fastest neutron, 
if neutrons were produced.  Neutrons with energy 50 MeV are possible to detect by the Chacaltaya detector, 
since the threshold energy is set at higher than $>$40 MeV.  
On the other hand, neutrons with the energy of 30 MeV arrive at the Earth after 22 minutes (1,300 seconds) 
later than the highest energy neutron.  The threshold energy of the Sierra Negra SNT was set at $>$30 MeV.  
However, the increase continued for 78 minutes.  Therefore it would be difficult to explain the signals observed at 
Mt. Sierra Negra by the direct hit of solar neutrons.

 The excess of the counting rate at Chacaltaya was $\sim$100 events/($m^2$minute) in the $>$40 MeV channel 
between 15:48 UT and 16:06 UT.  On the other hand, the S1 channel of Sierra Negra (the channel of charged particles 
with the energy higher than $>$30 MeV) showed 133 events / ($m^2$·minute) as the excess between 15:51-16:27 UT.  
The flux was observed with nearly the same intensity as that of Chacaltaya, and the excess start time was almost the same.  
However, the excess duration was completely different.  Therefore, in order to explain both enhancements consistently, 
we may introduce the assumption that the excess Chacaltaya was produced by the direct arrival of solar neutrons, 
while the excess Sierra Negra was produced by solar neutron decay protons (sndp). 

We may take into account another effect for the estimation of the flux of the sndp, since the energy of sndp 
is expected to be around a few GeV.  After transportation of the sndp in the magnetosphere, some of them 
may be trapped by the Earth$'$s magnetic field, but some of them will penetrate and arrive at Mt. Sierra Negra.  
The characteristic energy is called as the cut-off energy or Sp$ddot{o}$rer limit (rigidity) for low energy protons. 
The cut-off energy of Mt. Sierra Negra and Mt. Chacaltaya 
are estimated as 3$\sim$4 GeV and 11$\sim$12 GeV respectively (Shea and Smart, 2000).

The differential energy spectrum of protons near the cut-off energy has been measured in space 
by PAMELA (2009, 2016) and AMS (2000) detectors independently.  The result is shown in Figure 7 
by the green triangles for the cut-off energy of 3-4 GeV.  Therefore we will make an expected flux of sndp 
near the cut-off energy (the black circles in Figure 7), by multiplying the PAMELA$'$s observed differential flux 
of 3-4 GeV (the green triangles in Figure 7) (Casolino 2007, PAMELA 2009) by the expected neutron energy spectrum 
of En$^{-4}$dEn (the red diamond in Figure 7).   The energy spectrum of sndp 
beyond the cut-off energy is expected to reflect the neutron energy spectrum.  However, in the low energy region 
less than the cut-off energy, the energy spectrum of sndp is predicted to have almost constant value 
to the neutron spectrum of En$^{-4}$dEn.  The magnetic latitude of the observatory is estimated as 30$^{\circ}$N.  
The difference between the geographical latitude arises from the location of the magnetic-pole located at 
northern Canada (79$^{\circ}$N) over$ \sim$260$^{\circ}$E line (100$^{\circ}$W line) in 2001.  
  
(c) The increases observed by the GOES 11 satellite during 15:50-16:00 UT (Figure 3) was also produced 
by the neutron decay protons. They were decay products of high energy neutrons in the energy range
 between 80 and 400 MeV.  If these neutrons were produced at 15:47 UT, from the observed time, 
the parent neutrons had the energy between 80 and 400 MeV.  

It may be interesting to know that the flux of neutron decay protons differs two geostationary satellites.  
The increase was not observed in the detector onboard the GOES 10 satellite.  As for the longitude of both satellites on the 
Earth, the GOES 11 was situated at 114$^{\circ}$W, while the GOES 10 was located at 135$^{\circ}$W.  
In other words, the GOES 11 satellite was located in the right above American continent (just over the longitude of Mexico City).

(d) Another bump of electron and proton components around 16 UT was observed by the COSTEP instrument 
onboard the SOHO satellite (Figure 4).  They were also produced by the neutron decays in the outer magnetosphere.  
Now we focus on the neutron decay electrons.  Observation of electrons with energies higher than 2.64 MeV implies 
that initial neutrons should have a Lorentz factor $\gamma$=1.75 according to earlier predictions (Koi et al. 1993, Dorman 2010).  
It implies that original neutrons must have the kinetic energy higher than 700 MeV. 
 So initial protons must be accelerated beyond 1 GeV at the impulsive phase of the flare to produce high energy neutrons, 
En = 700 MeV.  The kinematics of neutron decay protons and electrons is given in Appendix 1.

\section{Flux Conversion observed by the ground level detectors to the top of atmosphere}

In order to compare each flux, in this section, we convert the flux measured by the ground level detectors to the top of the atmosphere.
 
a) ${\it  Deriving {\,}  flux{\,} of {\,}solar {\,}neutrons{\,} at{\,} the {\,}top{\,} of  {\,}atmosphere{\,} over{\,} Chacaltaya.} $

    In this section, let us derive the differential energy spectrum of solar neutrons at the top of the atmosphere.  
At 16 UT ( the local time at 12 LT), the atmospheric thickness over Chacaltaya is estimated as 550 g/$cm^2$.  
The differential energy spectrum is derived with use of one-minute value of the counting rate 
of the channel with E$>$40 MeV of the solar neutron detector.  
  In order to derive the differential energy spectrum, we introduce the hypothesis that neutrons 
were produced impulsively at the solar atmosphere. The flight time of neutrons depends on its kinetic energy.  
We fixed the production time at 15:47:00 UT.  
Then, the kinetic energy of neutrons is determined from the flight time (=arrival time $– $15:47:00 UT), 
so that the differential flux of solar neutrons is derived as a function of energy.  
The observed energy spectrum is given in Figure 8(a).

  After we derived the observed energy spectrum of solar neutrons by the detector, 
we converted it to the flux at the top of the atmosphere.  For this purpose we used two correction factors, 
the detection efficiency of solar neutrons by the SNT (Watanabe 2005) and 
the neutron attenuation curve in the atmosphere (Shibata 1994).  Combining these two correction factors 
into one curve, a correction curve is made.  The results is shown in Figure 8(b) as a function of incoming neutron energy.  

  Dividing the energy spectrum observed at Mt. Chacaltaya (Figure 8a) by the correction curve (Figure 8b), 
finally the differential energy spectrum of solar neutrons at the top of the atmosphere of Chacaltaya 
has been derived. The differential energy spectrum at the top of the atmosphere is presented in Figure 8(c).  
The differential flux has been already normalized to the flux per unit area ($/m^2$).

b) $\it Flux{\,} of{\,} solar {\,}neutron{\,} decay {\,}protons{\,} over {\,}at{\,} Sierra {\,}Negra$

Here we estimate the flux of the sndp over Mt. Sierra Negra.  We use the observed results in space by PAMELA (2007, 2009) 
and AMS (2000).  These results show that cosmic rays less energy than the cut-off energies have been observed.  
The green points of Figure 7 represent such effect.  The events recorded by the Mt. Sierra Negra 
Solar Neutron Telescope between 15:51 UT and 15:54 UT may correspond to the highest 
of the solar neutron decay protons (sndp).  Therefore, they may carry information on the flux 
near the cut-off energy of protons around E$_{cut}$$\sim$3 GeV.  
Incoming protons with energy about $\sim$3 GeV make nuclear interactions with air nuclei.  
As a result, neutral pions are produced with an energy around $\sim$1 GeV.  
These neutral pions immediately decay into two gamma rays and the gamma rays 
will make the electromagnetic cascade shower in the air.  The tails of the cascade shower 
enter into the scintillator and gamma rays involved in the shower are converted into electrons and positrons.  
From the intensity of electron and positron signals, the arrival intensity of protons over Mt. Sierra Negra may be estimated. 

Figure 7 (the black circles) suggests that the intensity of the sndp between Ep=1-3 GeV may be observed 
with an equal intensity and the spectrum is estimated to be flat.   Therefore, we choose energy bin 
width ($\Delta$E) of 3000 MeV (=4000-1000 MeV). Then we are able to estimate the differential flux of the sndp 
at 1-4 GeV as (4.4$\pm$3.0) / ($m^2$.min.MeV) at the top of the atmosphere.  
  On the other hand the flux of solar neutrons at the top of the atmosphere over Mt. Chacaltaya 
is estimated as 12.8, 1.8, and 0.5 events/($m^2$.min.MeV) at En=1, 2 and 3 GeV respectively.  
If we compare these numbers, the estimated solar neutron decay proton flux over Mt. Sierra Negra 
shows the Chacaltaya neutron flux. Given the assumptions made, we could say 
that the agreement is fairly good.  Each flux is summarized in Table 1.

  We are now preparing end to end simulation based on the GEANT4 code.  In this simulation, 
The primary energy spectrum measured by the PAMELA detector and the attenuation of gamma rays 
in the atmosphere will be taken into account. We expect that with these simulations the error bars 
on the flux of sndp will be reduced. Furthermore if you look carefully to Figure 2, you may notice 
another enhancement: the counting rate of the anti-counter registering charged particles 
with energy higher than 30 MeV, and the lower detector located under the scintillator (L1 channel)
 indicate an increase from 18:00 UT (14 LT in Bolivia and 12 LT in Mexico).  It is possible that protons 
were further accelerated into high energies by the shock acceleration mechanism (Tsuneta and Naito 1998). 
We will discuss this matter in the next paper, together with the results obtained by the GEANT4 simulation.

\section{ Production place of solar neutron decay protons}

  Before we discuss the production place of sndp in the interplanetary space, let us summarize 
the general situation of the interplanetary space around 16UT on November 7th 2004.  
Around the flare time, a very short gamma ray burst was observed by the gamma ray burst monitors 
onboard INTEGRAL and WIND satellite (INTEGRAL web site, WIND web site). 
According to INTEGRAL SPIACS data, a very short gamma ray burst was recorded at 15:49:30 UT.  
The KONUS detector onboard the WIND satellite also detected the short gamma ray signal and they 
detected 2.2 MeV line gamma rays.  The 2.2 MeV line gamma rays are emitted when a neutron
 is captured by a proton and to form a deuterium.  The RHESSI satellite detected hard X-rays after 16:05 UT. 
{\,} In Figure 9, we present two images of the flare at 15:36 UT and 15:48 UT.  
The images were taken by an ultraviolet telescope onboard SOHO.  
Within 16 minutes, an arch was emphasized in the image at 15:48 UT. 
 It is shown by the white artificial arc in the left side of Figure 9.  
Particles may possibly be accelerated within this arc.

Now let us describe briefly the characteristics of the CME1. According to the magnetometer measurement 
onboard ACE satellite (ACE web site) that stays at the Lagrange point L1, the maximum field strength of the CME1 was 40 nT.  
This first CME1 (on the right side in Figure 6) was produced by the M9.3 flare of November 6th, not by the X2.0 flare of November 7th.  
According to the IPS observation, the CME2 arrived on the Earth around 9UT of November 9th (Tokumaru 2013). 
The CME2 produced at 16UT is depicted near the Sun in Figure 6. 
The CME1 was already expanded in the interplanetary space at 16 UT on November 7.  
From the record of the ACE satellite, the diameter of the CME1 may be estimated. 
 We estimated it as $\ell$1= $10^7$ km.  The width of the CME1 was calculated by multiplying
 the time (9,000 sec) by speed (1,150 km/sec).   Then we can estimate the maximum momentum 
of charged particles that will be trapped inside the CME1.  By putting 
the numerical values into a simple equation pc=300 H$\rho$, as H=4$\times$10$^{-4 }$gauss, $\rho$= 0.5 $\times$10$^{12 }$cm, 
we get pc=6$\times10^{10}$ eV=60 GeV.  The value is quite high.  In other words, low energy charged particles 
may be trapped by the “CME barrier”, so that neutron decay protons produced between the Sun and the CME1 
could not arrive the detector located near the Earth.  Therefore, the observed sndp must be produced 
in between the front side of CME1 and the Earth. This region is pictorially shown in Figure 6 as D1.  
(Drawing is not proportional to the actual distance. ) The direct line approximation of the decay path
may be guaranteed by the preceding calculations, at least during 60 minutes (Sakai and Muraki 1993, Sakai et al. 1997).

   We examine “another barrier” for charged particles.  It is the wall produced by the magnetosphere.  
The length of the magnetic-sheath $\ell$3 is estimated as  $\ell$3= 2.7$\times$$R_{earth}$ = 17,000 km = 1.7$ \times$10$^{9} $cm.  
The field strength is estimated as between 20 nT $–$ 40 nT.  Therefore, we choose 30 nT.  
Again, putting these values into the equation of pc=300 H$\rho$, we will get pc= 15 MeV.   
This time, the threshold energy is quite low, however protons with the energy less than 15 MeV
 produced between the CME1 and the bow shock cannot penetrate inside the magnetosphere.  
In present case, the initial neutrons observed by the GOES detector had an energy between 80 $-$ 400  MeV.  
Therefore the decay products (sndp) between Ep=80 MeV and 400 MeV can penetrate the magneto-sheath. 
 (For the reference, we provide the length of $\ell$2 and $\ell$4 as l2=$10^7$ km =0.066 AU, and $\ell$4= 56,700 km respectively.  
The flight length from the front edge of the CME1 to the mountain detector D1 
is estimated as D1= 1$\times10^7 $km + 67,000 km$\approx$1$\times$10$^{7}$ km.)  

   In the events observed in 1982 June 3, the sndp were registered for almost 12 hours (Evenson et al. 1983, 1990), 
while in the present event the excess was observed only during 75 minutes.  The difference may be 
due to the existence of the preceding CME (CME1).  In the 1982 June 3rd flare, 
there was no preceding CME (Solar Geophysical Data website).   
However in the event of 1989 October 19th, the sndp were observed for 30 minutes and the signals of sndp
 were masked by the strong accelerated proton beam produced by the X13 flare (Shea et al. 1991).  
                                               
\section{Conclusive Remarks}

We have here presented evidence corresponding to registers in two spacecrafts and two ground based detectors. 
Taken together, observations admit an interpretation that would be consistent with the observation of solar neutron decay protons.
Solar neutron decay protons were first reported in 1981 by Paul Evenson, Peter Meyer, Roger Pyle. 
 Ruffolo and other researchers discussed solar neutron decay electrons and protons afterwards (Ruffolo 1991, Dr$\ddot{o}$ge et al. 1996).  
However, we do not know any report of this kind of events after that of 19, October 1989.  
To the best of our knowledge, there are only three early reports on the detection of neutron decay protons and electrons 
onboard satellites and one report of neutron decay protons by ground level detectors in October 19th, 1989.  
The event of October 1989, was also discussed (Koi et al. 1993).  So, present event may be the second case 
where neutron decay protons were registered by a ground-based detector.  Further study is necessary 
to understand this phenomenon deeply.  However, we wanted to publish this quick report to call community attention 
in the search of neutron decay protons and electrons.

\section{Competing interests}

The authors declare that they have no competing interests. 

\section{Author contributions}

YM, JFVG, YMa, SS, TSako, and TS have constructed the Mt. Sierra Negra Solar Neutron Telescope, 
while YMa, TSako, YM and PM constructed the Chacaltaya solar neutron detector.  
YM, JFVG, EO and SS have made the data analysis.  
SM and MT provided the solar image and solar wind data respectively. SS prepared Appendix 1.  
All members jointed discussions.

\section{Acknowledgments}

The authors acknowledge INAOE personnel and authorities for their continued support 
in providing the services needed to keep the Mt. Sierra Negra SNT functioning well. 
We also acknowledge UNAM-PA`IIT partial support through grant IN-104115. 
The authors express the acknowledgment to the staffs of physics department of UMSA, 
who have kept the Solar Neutron Detector located at Mt. Chacaltaya under good condition for 28 years. 
We express sincere thanks to the satellite teams; particularly SOHO (LASCO, COSTEP, 
and EIT instruments) and ACE satellites, who provided valuable data available from the web site. 
This work has been carried out, being based on the international solar neutron telescope (SONTEL) network 
data reserved in the storage system of ISEE, Nagoya University.

\section{Author details}

1) Institute for Space Earth Environmental Research, Nagoya University, Nagoya 464-8601, Japan, 
2) Instituto de Geofisica, UNAM, 04510, Mexico D. F., Mexico, 
3) Escuela Nacional de Ciencias de la Tierra, UNAM. Ciudad de Mexico. 04510. Mexico, 
4) Engineering Science Laboratory, Chubu University, Kasugai, Aichi 487-0027, Japan, 
5) Astronomy Observatory, Chubu University, Kasugai, Aichi 487-0027, Japan, 
6) Institute for Cosmic Ray Research, The University of Tokyo, Kashiwa, Chiba 277-8582, Japan,  
7) Physical Science laboratory, College of Industrial Technologies, Nihon University, Narashino, Chiba 275-0006, Japan, 
8) Department of Information Science, Yamanashi Gakuin University, Kofu 400-8575, Yamanashi, Japan, and 
9) Department of Physics, UMSA, La Paz, Bolivia

\section{References}

$\ast$) ACE web site;\\
 http:// www.srl.caltech.edu/ACE/ASC/DATA/level3/mag/ACECpec.cgi?LATEST=1\\
$\ast$) Alex A. (2018) Script Lecture PHY 432, Physics with muons: from atomic physics to Solid state physics (psi ch.) (Access is available from website)\\
$\ast$) AMS collaboration; Alearaz et al. (2000) Protons in near earth orbits. Physics Letters B 472:215-226  https://doi.org/10.1016/S0370-2693(99)01427-6\\
$\ast$) Burman R.L. and Smith E.S (1989) LA-11502-MS, Parametrization of Pion production and Reaction cross-section at LAMPF energies (Access is available from website.)\\
$\ast$) Casolino M et al. (2007) Two Years of Flight of the Pamela Experiment: Results and Perspectives. J. Phys. Soc. Jpn. Suppl. A 78:35-40, 
M. Casolino et al. (2007) Observation of primary, trapped and quasi trapped particles with PAMELA experiment. Proceed. 30th ICRC (Merida) 1:709-712, 
Also in the rapporteur talk by E.O. Flückiger (2007) Ground Level Events and Terrestrial Effects.  Proceed. 30th ICRC (Merida) 6: 239-253\\
$\ast$) Dorman LI, Valdés-Galicia JF, Dorman VI (1999) Numerical simulation and analytical description of solar neutron transport in the Earth's atmosphere, Journal Geophysical Research 104 (A10): 22417-22426. Doi.org/10.1029/1999jA900182\\
$\ast$) Dorman LI, Dorman IV, and Valdés-Galicia JF (1997) Simulation of solar neutron scattering and attenuation in the Earth’s atmosphere for different initial zenith angles. Proceed. 25th ICRC (Durban) 1:25-28\\
$\ast$) Dorman L (2010) Solar Neutrons and Related Phenomena (Springer). pp 374-377doi: 10.1007/978-90-481-3737-4\\
$\ast$) Droge W, Ruffolo D, and Klecker B (1996) Observation of electrons from the decay of solar flare neutrons.  ApJ 464: L87-L90\\
$\ast$) Evenson P, Meyer P, and Roger Pyle K (1983) Protons from the decay of solar flare neutrons.  ApJ 274:875-882
$\ast$) Evenson P, Kroger R, Meyer P, and Reames D (1990) Solar Neutron Decay Proton Observations in Cycle 21.  ApJ Suppl. 73:273-277\\
$\ast$) INTEGRAL web site: https://www.isdc.unige.ch/integral/science/grb$\sharp$ACS \\
$\ast$) Kamiya K, Koga K. Matsumoto H., Masuda S, Muraki Y., Tajima H, and Shibata S (2019) Solar Neutrons observed from September 4 to 10 2017 by SEDA-FIB, PoS (ICRC2019) 1150.\\
$\ast$) Koi T. et al.(1993) Prediction of electrons as decay products of solar neutrons. Proceeding of 23rd ICRC (Calgary) 3:151-154\\
$\ast$) Muraki Y., Matsubara Y., Masuda S., Sakakibara S., Sako T., Watanabe K., Bütikofer R., Flückiger E.O., Chilingarian A., Hovsepyan G., Kakimoto F., Terasawa T., Tsunesada Y., Tokuno H., Velarde A., Evenson P., Poirier J., and Sakai T. (2008)  Detection of high-energy solar neutrons and protons by ground level detectors on April 15, 2001. Astroparticle Physics 29:229-242  Doi:10.1016/j.astropartphys.2007.12.007\\
$\ast$) Muraki Y. Valdés-Galicia JF, González LX, Kamiya K, Katayose Y, Koga K, Matsumoto H, Masuda S, Matsubara Y, Nagai Y, Ohnishi M, Ozawa S, Sako T, Shibata, S, Takita M, Tanaka Y, Tsuchiya H, Watanabe K, and Zhang JL (2020) Possible detection of solar gamma-rays by ground-level detectors in solar flares on 2011 March 7.  Pub. Astron. Soc. Japan 72:1-17  Doi: 10.1093/pasj/psz141\\
$ast$) PAMELA collaboration; N. De Simone et al. (2009) Comparison of models and measurements of protons of trapped and secondary origin with PAMELA experiment. Proceed. 31st ICRC (Lodz) icrc0795 \\
Adriani et al. (2016) PAMELA's measurements of geomagnetic cutoff variations during the 14 December 2006 storm. \\
Space Weather 14 210-220     doi:10.1002/2016SW001364 \\
$\ast$) Ruffolo D (1991) Interplanetary transport of decay protons from solar flare neutrons.  Astrophysical Journal 388:688-698 Doi: 10.1086/170756\\
$\ast$) Sakai T. and Muraki Y. (1993) Solar neutron Decay Proton. Proceed of 23rd ICRC (Calgary) 3:147S.  Website: articles.ads.abc.harvard.edu/pdf/1993ICRC….3..1475\\
$\ast$) Sakai T, Kato M, and Muraki Y (1997) Propagation of Solar neutron decay protons near the Earth. Geomag. Geoelectr. 49:1105-1113\\
$\ast$) Shea MA, Smart DF, Wilson MD, and Flückiger EO (1991) Possible ground-level measurement of solar neutron decay protons during the 19 October 1989 solar cosmic ray event.  Geophys. Res. Lett. 18 : 829-832.  Doi.org/10.1029/90GL02668\\
$\ast$) Shea MA and Smart DF (2000) Cosmic Ray Implications for Human Health. Space Science Review 93:187-205\\
$\ast$) Shibata S (1994) Propagation of solar neutrons through the atmosphere of the Earth. Journal of Geophysical Research 99: 6651-6665\\
$\ast$) SOHO CME catalog website: https://cdawweb.gsfc.nasa.gov\\
$\ast$)SOHO-COSTEP-EPHIN : http://www2.physik.unikiel.de/SOHO/phpeph/EPHIN.htm\\
$\ast$) Solar Geophysical Data web site: \\ https://www.ngdc.noaa.gov/stp\\
/space-weather/online-publications/stp{\,}sgd/1982/sgd8212c.pdf \\
$\ast$) Tsuchiya H.et al. (1999) Detection efficiency of new solar neutron detector.  Proceed. 26th ICRC (Salt Lake City) 7: 363\\
$\ast$) Tsuchiya H et al. (2001) Arrival of solar neutrons from large zenith angle.  Proceed. 27th ICRC (Hamburg) 8: 3056-3059\\
$\ast$) Tsuneta S. and Naito T (1998) Fermi acceleration at the fast shock in a solar flare and the inclusive loop-top hard X-ray source. APJL 67:495\\
$\ast$) Tokumaru, M (2013) Three-dimensional exploration of the solar wind using observations of interplanetary scintillation, Proceedings of the Japan Academy Ser. B, 89(2):67-79   doi:10.2183/pjab.89.67.\\
$\ast$) Valdés-Galicia JF, Muraki Y., Sako T, Musalem O, Huertado A, Gonzalez X, Matsubara Y, Watanabe K, Hirano N, Tateiwa N, Shibata S, and Sakai T (2004) An improved solar neutron telescope installed at a very high altitude in México. NIMA 535: 656-664  doi.org/10.1016/j.nima.2004.06.148\\
$\ast$) Valdés-Galicia JF, Muraki Y,  Watanabe K, Matsubara Y, Sako T,  Gonzalez X, Musalem O, Huertado A (2009) Solar neutron events as a tool to study particle acceleration at the Sun, Advances in Space Research, 43: 565-572. https://doi.org/10.1016/j.asr2008.09.023 \\
$\ast$) Watanabe K. (2005) Proceeding of the Cosmic-ray Research Section of Nagoya University 46(2), pp1-249.  Solar neutron Events associated with Large Solar Flares in Solar Cycle 23 (PhD thesis in English)\\
$\ast$) WIND web site:  https://asd.gsfc.nasa.gov\\
%
%

\section{Appendix 1}

Appendix 1 ---kinematics of neutron decay electrons ---

Let us define the parameters that are used
for the estimation of the highest energy of electrons from neutrons.
Here $m_{n}$, $m_{p}$, and $m_{e}$ represents
the mass of neutron, proton, and electron respectively.
$E_{p}^{*}$ and $E_{e}^{*}$ are the energy of proton and electron
in the center of momentum system (in the neutron rest frame).
In this calculation, we take the neutrino mass zero.
Now we study an extreme case;
proton and neutrino are emitted backward,
while electron is emitted forward.
Then the electron energy in the center of momentum system is expressed by
$$
E_{e}^{*}=((m_{n}c^{2}-m_{p}c^{2})^2+me^{2}c^{4})/(2(m_{n}c^{2}-m_{p}c^{2})).
\qquad (1)
$$
We know $(m_{n}c^2-m_{p}c^2)=1.29\,{\rm MeV}$
and
$m_{e}c^2=0.511\,{\rm MeV}$.           
Putting these values in Equation (1),
we get $E_{e}^{*}=0.746\,{\rm MeV}$.\\

From a relation, $(P_{e}^{*}c)^2=E_{e}^{*2}-m^{2}c^{4}$,
and
$P_{e}^{*}c=0.544\,{\rm MeV}$.\\
  
The Lorentz transform is expressed as
$$
E_{e}=\gamma(E_{e}^{*}+\gamma\beta P_{e}^{*}c \cos{{\theta}^{*}}),
\qquad (2)
$$
where $\gamma$ represents the Lorentz factor of incident neutron.\\
So the electron energy in the laboratory frame Ee can be expressed by
$$
E_{e}=\gamma(0.746+\gamma\beta\times 0.544)\,{\rm MeV}\ \
{\rm in\ case}\ \cos{{\theta}^{*}}=1.
\qquad (3)
$$

The threshold energy of COSTEP detector onboard SOHO satellite is 2.64 MeV.
Therefore solving the equation,
we will get the minimum Lorentz factor of solar neutrons($\gamma$) by
$$
2.64=\gamma(0.746+\gamma\beta\times 0.554).
\qquad (4)
$$                                                                                               
When we put $\gamma=1.75$ and $\beta=0.821$ in Eq.(4),
we will get $E_{e}\approx 2.67\,{\rm MeV}$.\\

Let us provide another extreme case,
{\it i.e.},
the proton is emitted forward
and the electron and neutrino are emitted backward.
The maximum energy of the proton in the center of momentum system
is expressed by
$$
E_{p}^{*}=(m_{n}^{2}c^{4}+m_{p}^{2}c^{4}-m_{e}^{2}c^{4})/2m_{n}c^2.
\qquad (5)
$$

Since $E_{p}^{*}=m_{p}c^{2}$ and $P_{p}^{*}\approx 0$, in Eq.(2),
we put $E_{p}^{*}$ instead of $E_{e}^{*}$,\\
then we get
$$
E_{p}=\gamma m_{p}c^{2}=(m_{p}/m_{n})E_{n}.
\qquad (6)
$$
Therefore the proton emitted the forward direction
takes almost all the energy (0.999) of the neutron.\\

\begin{figure}[ht] 
\hspace{-20mm}
  \includegraphics[width=160mm]{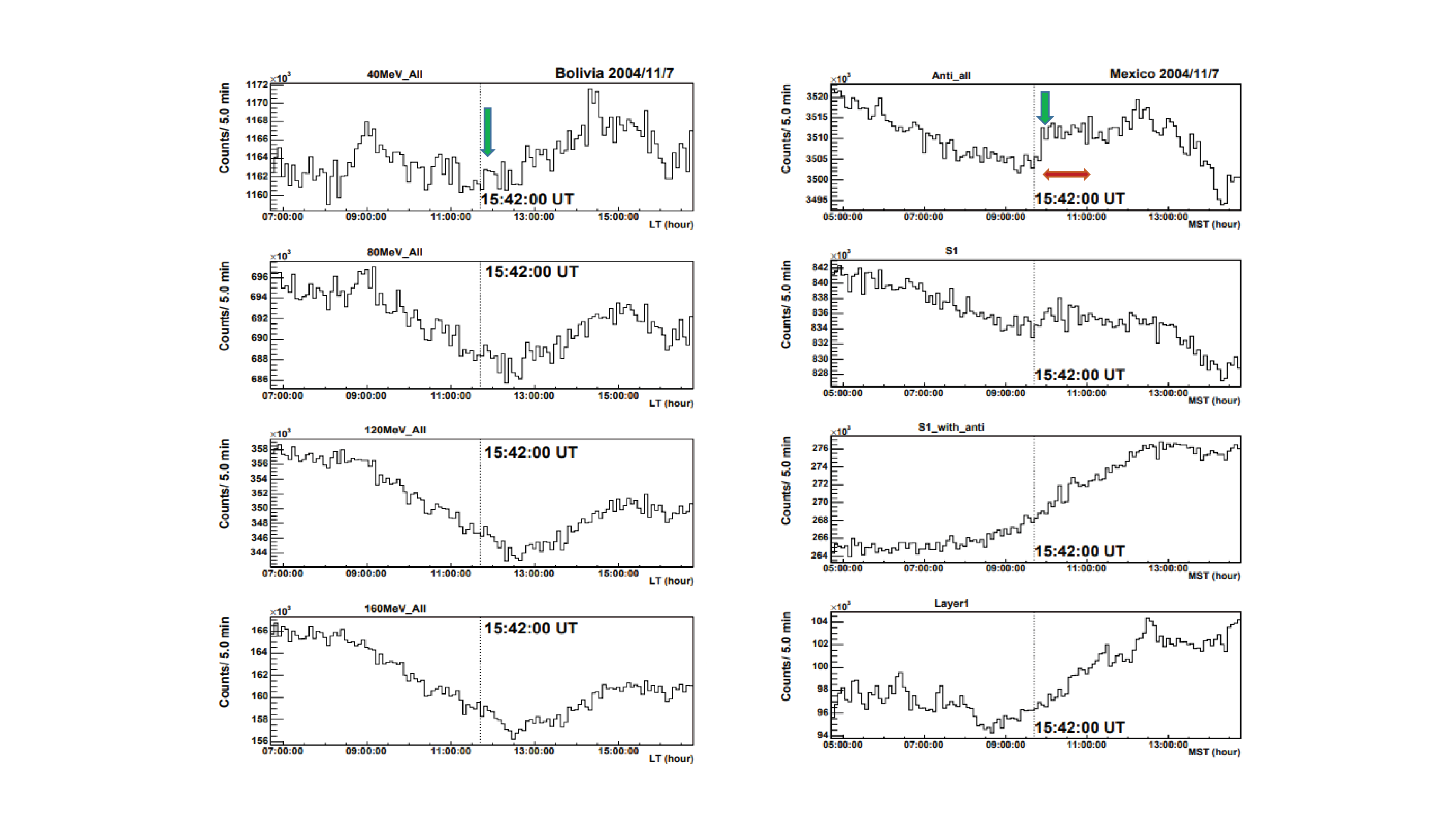} 
  \caption{ (left). The 5-minute-value of the counting rate of the solar neutron detector located at Mt. Chacaltaya in Bolivia (5,250m a.s.l.).  
From top to bottom, it presents the counting rate of charged particles with the deposit energy 
higher than $>$40, $>$80, $>$120, and $>$160 MeV respectively.  The threshold energy of the detector
 is calibrated by using the deposit energy of the minimum ionizing particles like muons.  
The arrow corresponds to the time of the enhanced counting rate, around 16 UT and 
the vertical dotted line represents the GOES flare start time. The horizontal presents the local time. }
\caption{ (right). The 5-minute-value of the counting rate of the solar neutron telescope located at Mt. Sierra Negra in Mexico (4,600m a.s.l.). 
 From top to bottom, the picture corresponds to the counting rate of all charged particles, charged particles with energy higher than 30 MeV, 
neutral particles with energy higher than 30 MeV, and the proportional counter located underneath the scintillator.  
We call it L1 channel (Layer 1). The L1 channel is triggered with the signal of S1 ($>$30MeV).  
The L1 channel is used for the identification of neutrons and gamma rays.  
The vertical dotted line presents the GOES flare start time (UT) and the horizontal arrow indicates the time of the enhanced counting rate. } 
\end{figure} 

\begin{figure}[ht] 
\hspace{0mm}
\vspace{20mm}
  \includegraphics[width=180mm]{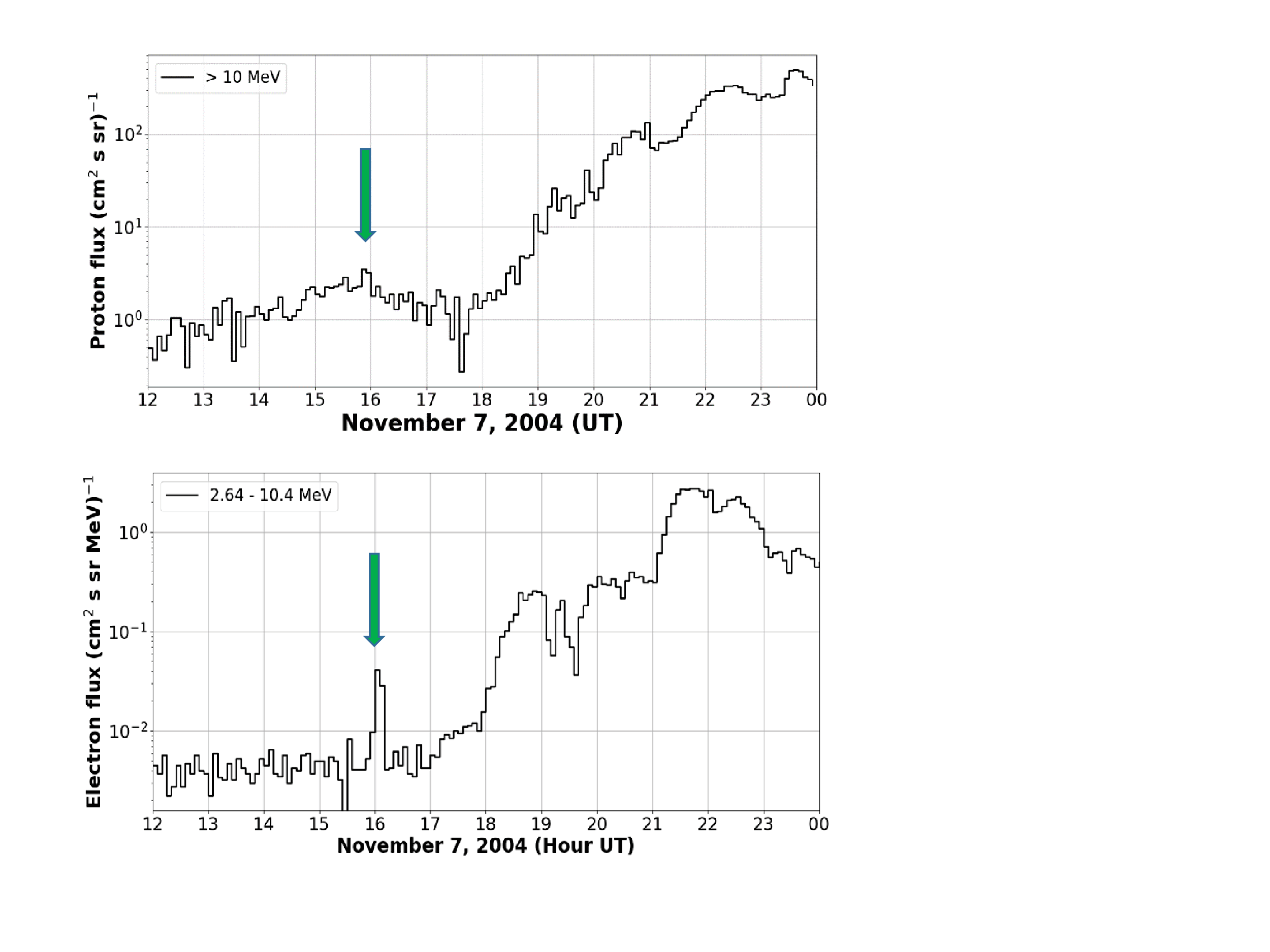} 
  \caption{(top). The time variation of the proton intensity with energy higher than 10 MeV measured by the proton counter onboard the GOES 11 satellite.  
An enhancement can be recognized 15:50-16:00 UT (indicated by an arrow).  The enhancement corresponds to the impulsive phase 
of the solar flare X2.0 and may be produced by the neutron decay protons. } 
\caption{(bottom)  The time variation of the electron intensity with the energy 2.64-10.4 MeV measured by the COSTEP detector onboard the SOHO satellite. 
The enhanced time is indicated by the arrow. If the enhancement was produced by solar neutron decay electrons, 
the initial neutron should have the Lorentz factor $\gamma$=1.75 according to Koi et al. (1993). }
\end{figure} 

\begin{figure}[ht] 
\hspace{-20mm}
\vspace{-20mm}
  \includegraphics[width=180mm]{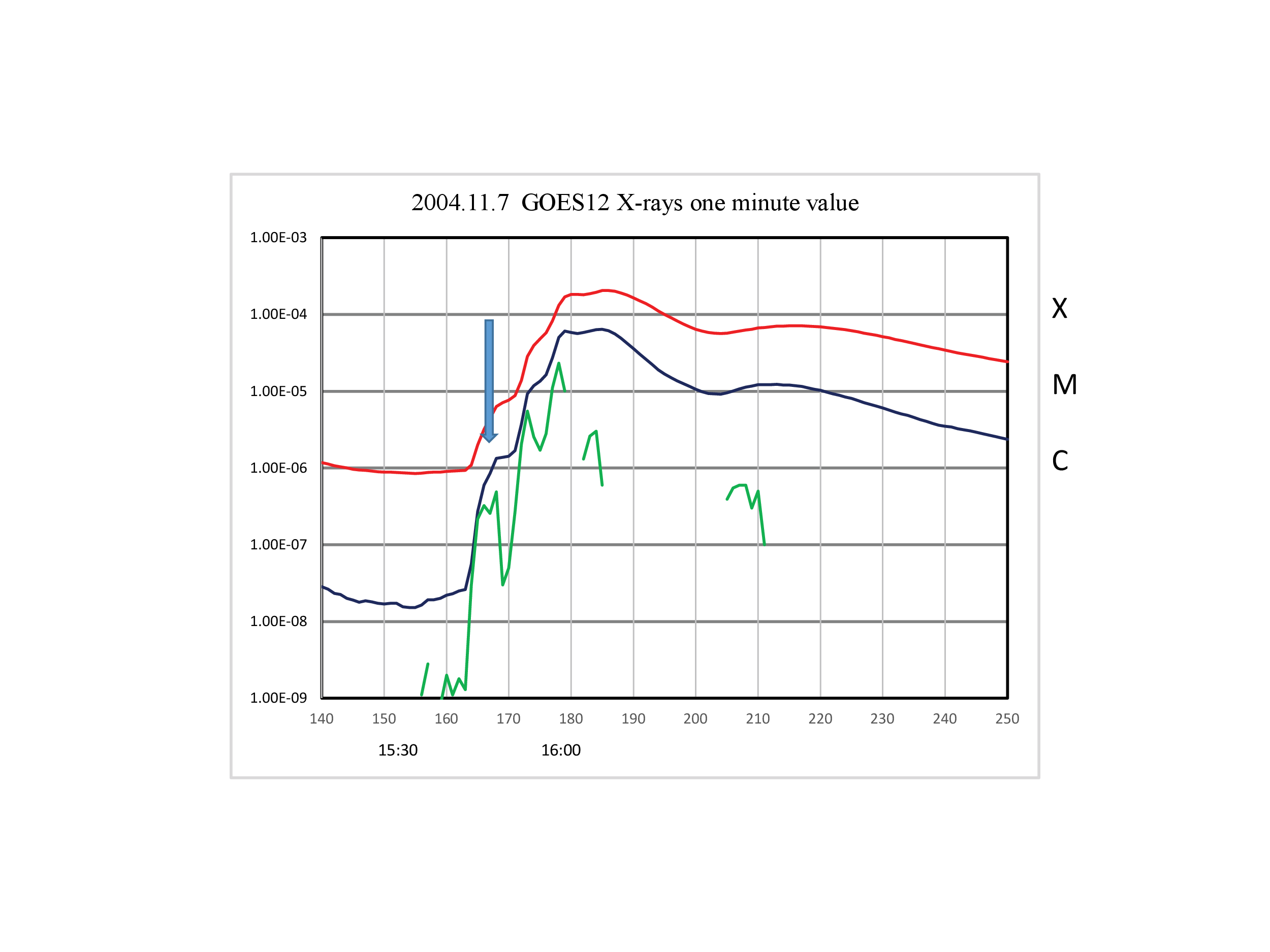} 
  \vspace{10mm}
  \caption{The one-minute time profile of GOES X-ray intensity from 15:20 UT to 
17:10 UT November 7, 2004.  The red line and the blue line correspond to the X-rays 
with the wave length 1-8${\AA}$ and 0.5-4${AA}$ respectively.  The green line represents the derivative of the short band of X-rays. 
We assume that neutrons were produced instantaneously at 15:47:00 UT, when there is a change of slope of the emission. } 
\end{figure} 

\begin{figure}[ht] 
\hspace{-10mm}
  \includegraphics[width=160mm]{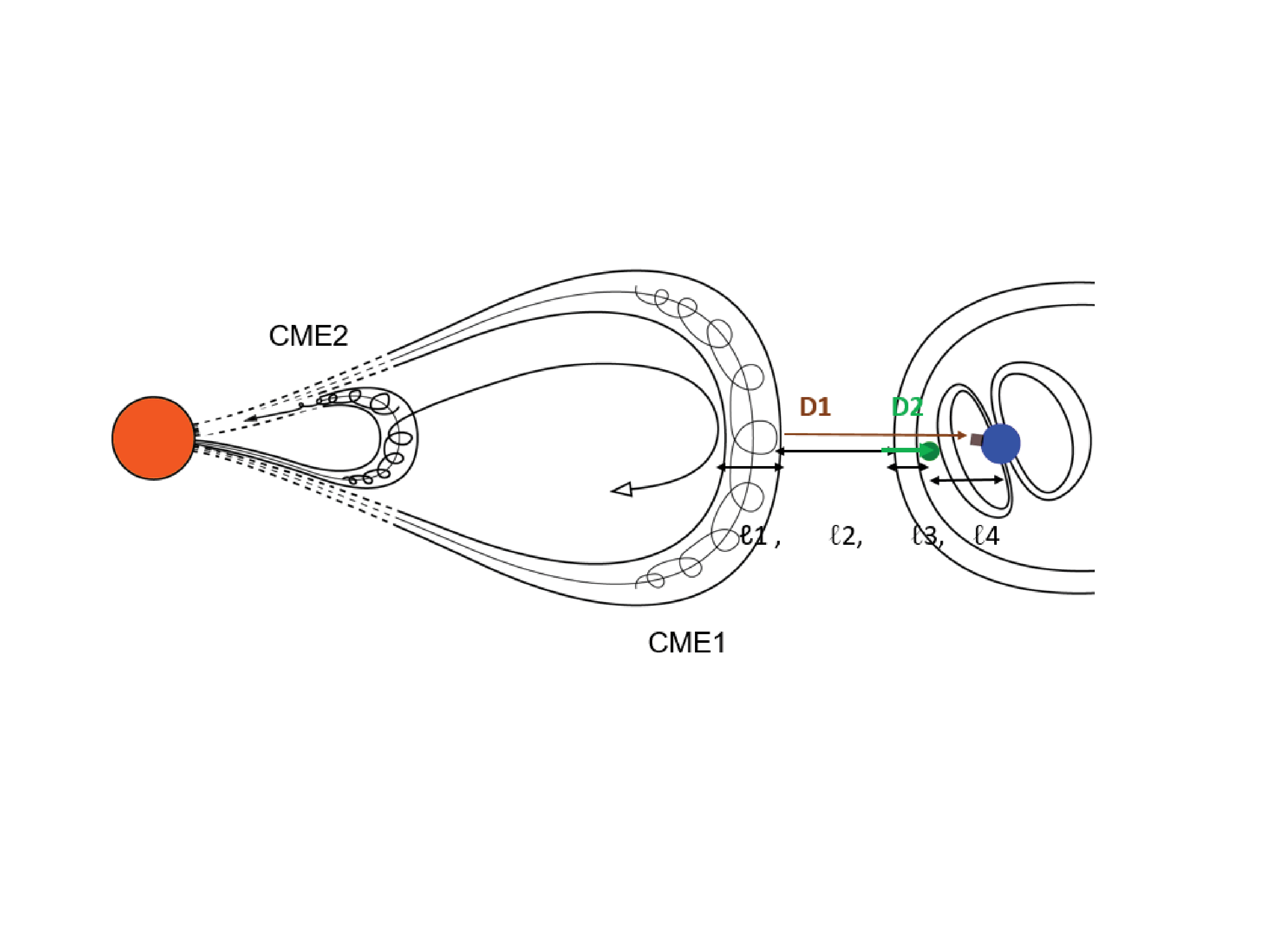} 
  \caption{The interplanetary plasma scenario around 16 UT of November 7th, 2004 is depicted. 
had not yet arrived near the Earth at 16 UT, situated about 1${\times}10^{7}$ km (${\ell}$2) away from the (${\ell}$3) of the Earth.  
The front of the magnetosphere was extended at about 9${R_earth}$ (${\ell}$4).   The green star presents the position of the GOES satellite 
and the brown mark corresponds to Mt. Sierra Negra observatory.  D1 represents the distance from the CME
front to the mountain laboratory, and D2 (=2.2${\times}10^{4}$ km) depicts the distance from the bow of the magneto pause to the GOES satellite} 
\end{figure} 

\begin{figure}[ht] 
\hspace{-10mm}
  \includegraphics[width=160mm]{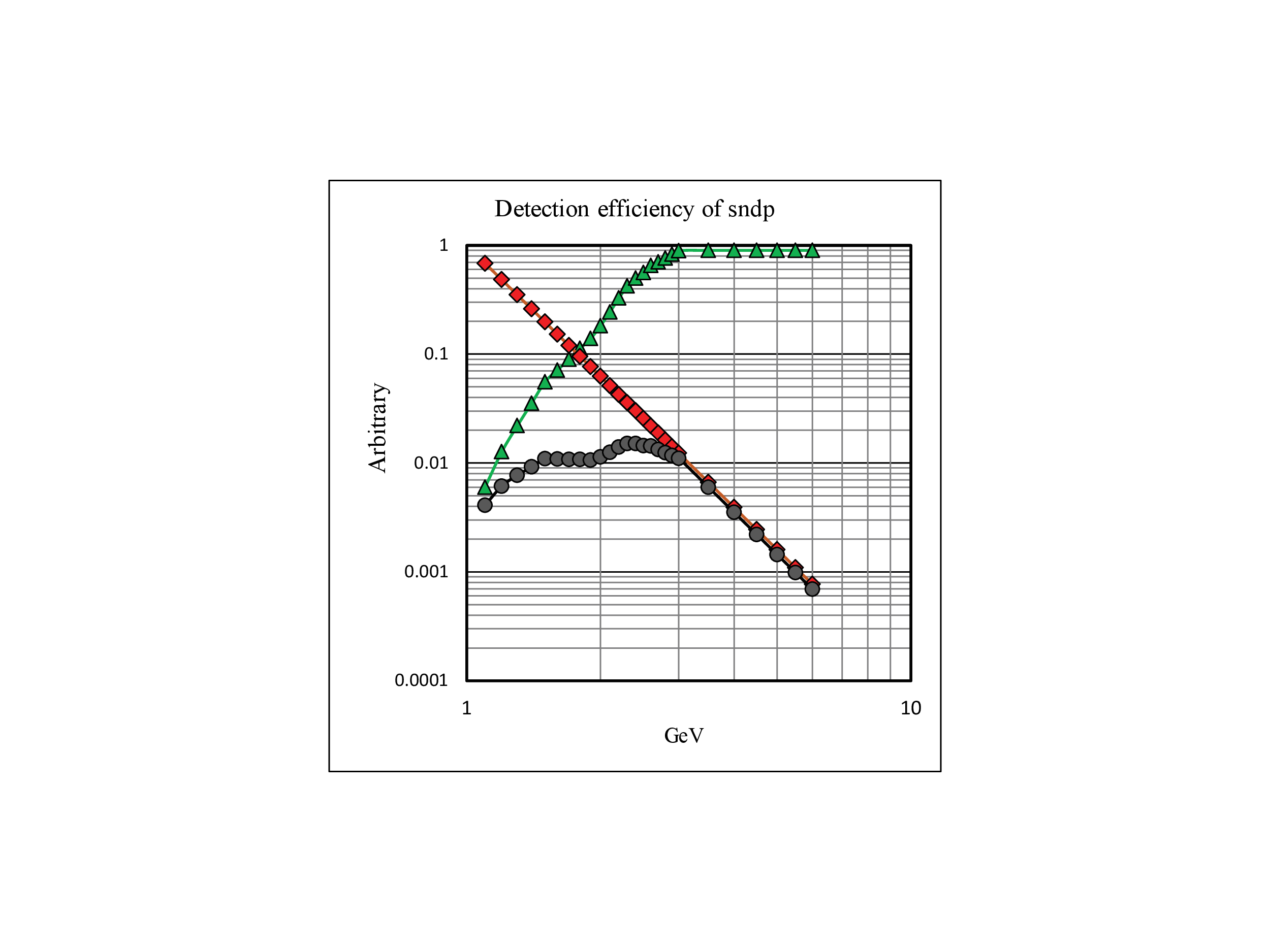} 
  \caption{The expected solar neutron decay proton (sndp) flux at Mt. Sierra Negra (the black circles). 
The red diamonds represent a possible production spectrum of solar neutrons for the energy spectrum of En$^{-4}$dEn.  
The green triangle presents the observed proton flux by the PAMELA detector near the cut-off energy of 3-4 GeV. 
 Beyond the cut-off energy, the energy spectrum of the sndp represents the neutron spectrum } 
\end{figure} 

\begin{figure}[ht] 
\hspace{-20mm}
  \includegraphics[width=160mm]{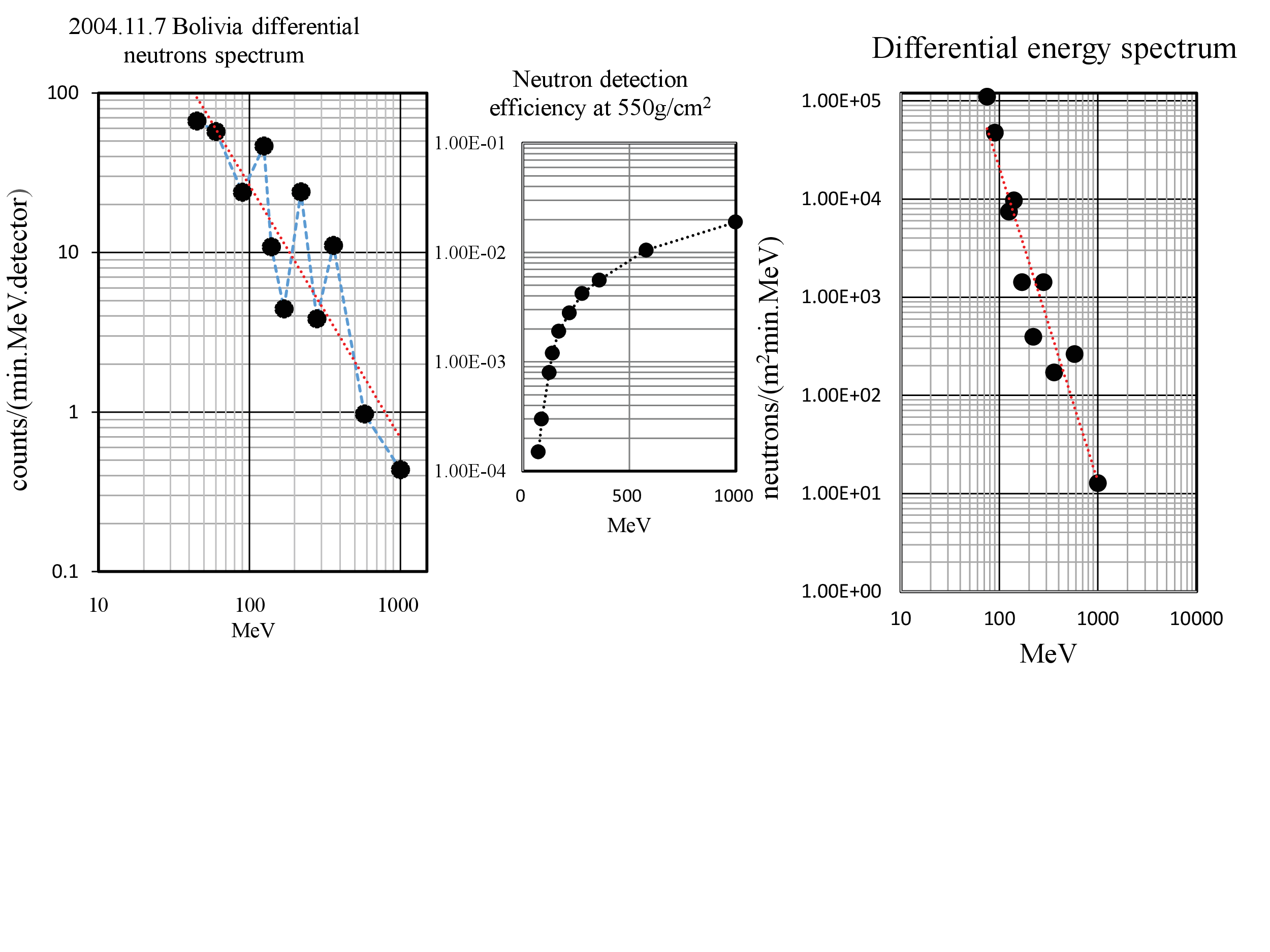} 
  \caption{(a) (left) the original energy spectrum observed at Mt. Chacaltaya, 
     (b) (center)  the correction factor for the energy spectrum to the observed spectrum.
     (c) (right) The estimated energy spectrum at the top of the atmosphere over Mt. Chacaltaya in units of flux/(cm$^{2}$·min. MeV).}  
\end{figure}

\begin{figure}[h] 
  \includegraphics[width=160mm]{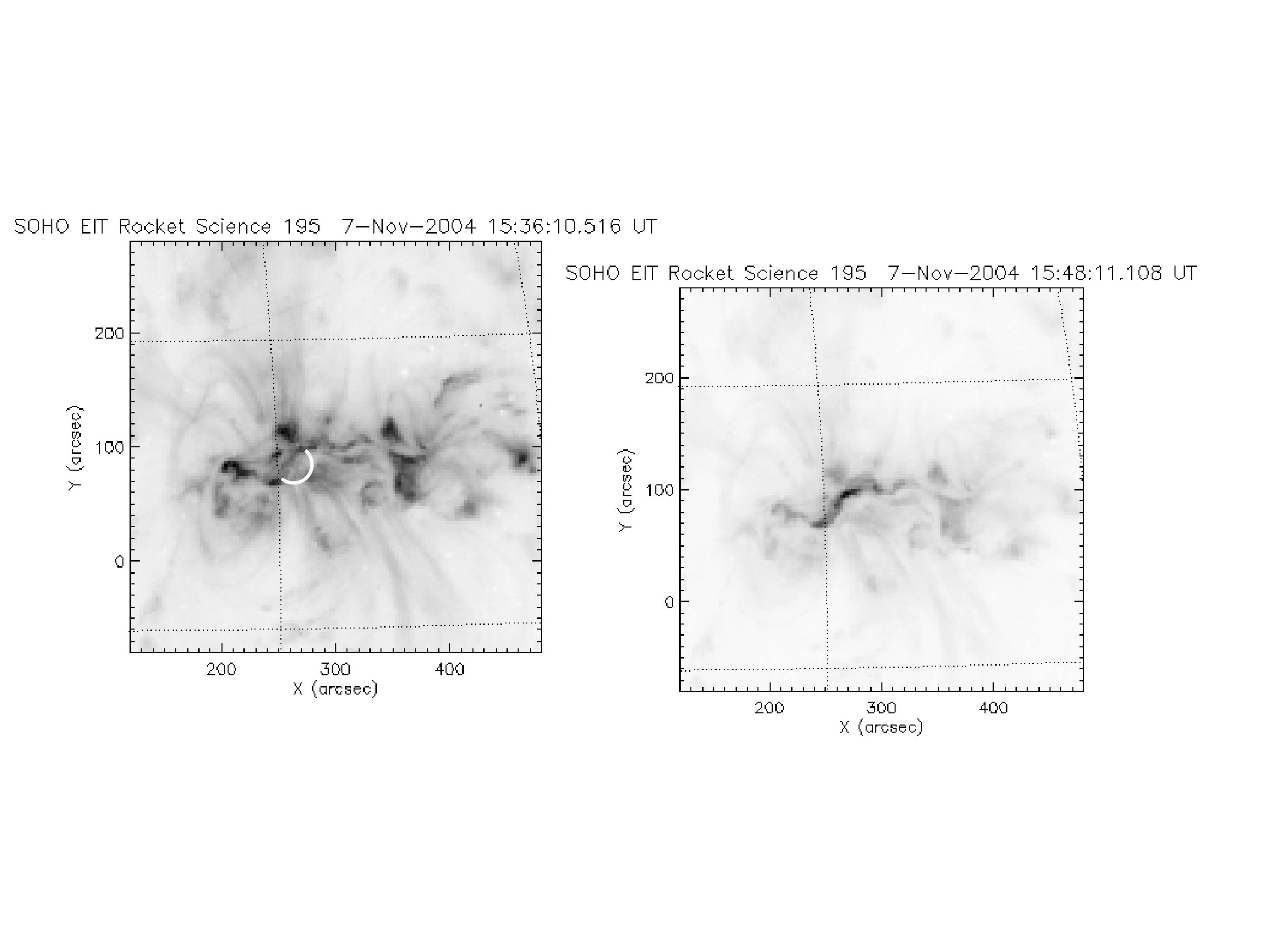} 
  \caption{ Solar images taken by the EIT telescope onboard SOHO satellite at 15:36 UT (left) and 15:48 UT (right).  
The UV wave length is 19.5 nm. We assume that neutrons were instantaneously produced at 17:47 UT.  
If you compare the two images (a) and (b), you will notice a plasma arcade in the image of  15:48 UT.  
The arc is shown by a white arc in the image of 15:36 UT and possibly it had an important role for the particle acceleration.} 
\end{figure}

\begin{figure}[ht] 
  \includegraphics[width=160mm]{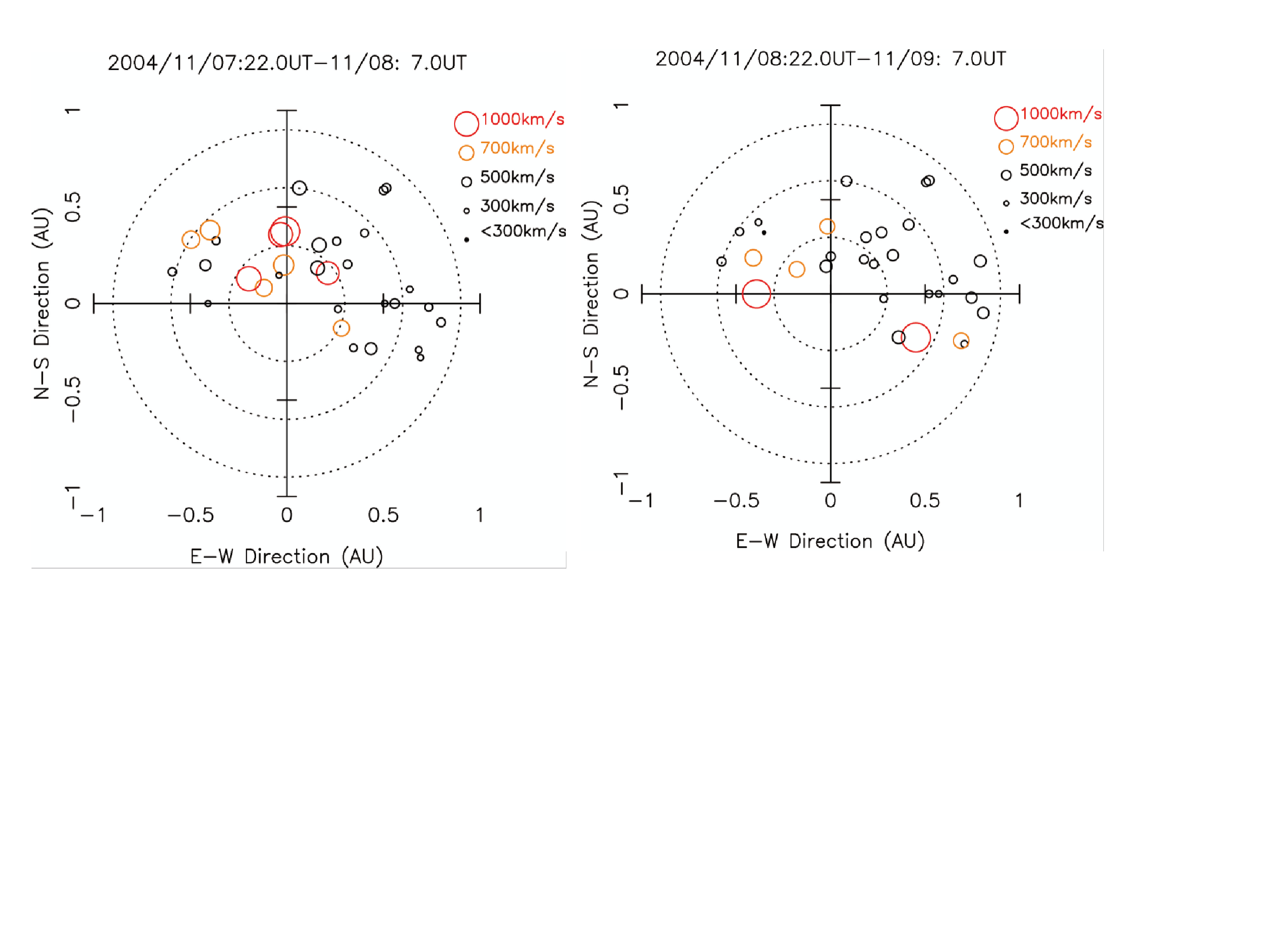} 
  \vspace{-40mm}
  \caption{The sky projection maps of the solar wind speed derived from the observations of the interplanetary scintillation device (IPS) 
using the 327-MHz multi-station system of Nagoya University (Tokumaru, 2013).  
The data were accumulated during the periods (left side) between 22 UT of November 7, 
and 7 UT of November 8, 2004, and (right side) between 22 UT of November 8 and 7 UT of November 9, 2004 respectively. 
 Four IPS antennas observed IPS for the same radio sources simultaneously. The solar wind speed was determined by 
detecting the time lag between the variations of the intensity at the separated stations. 
The center of the map corresponds to the location of the Sun, and circles indicate the relative positions of lines-of-sight for radio sources. 
The radius of the circle represents the solar wind speed. Red circles denote high speed ($>$1000 km/s) plasma flow. 
The solar offset distances of high-speed data for November 8-9 (right panel) are larger than those for November 7-8 (left panel). 
This fact suggests that the expanding CME was approaching near the Earth.  T
he shock in association with the X2.0 flare arrived on Earth at ~09 UT in November 9th} 
\end{figure}

\begin{figure}[ht] 
  \includegraphics[width=120mm]{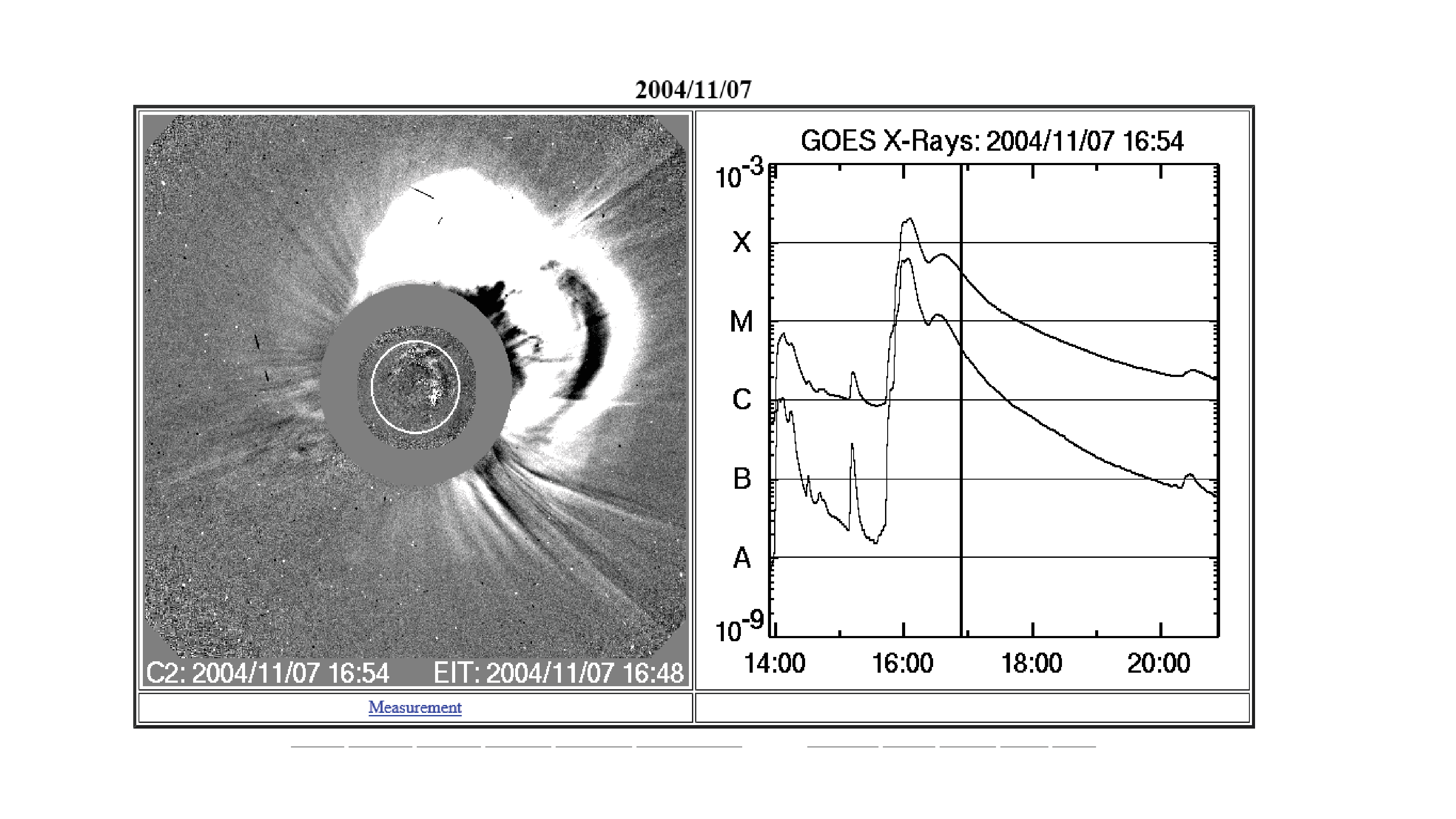} 
  \caption{ The image of the CME-CME interaction taken by the SOHO LASCO telescope at 16:54 UT on November 7th, 2004. 
The central white circle represents the solar surface.} 
\end{figure} 
\begin{figure}[ht] 
  \includegraphics[width=120mm]{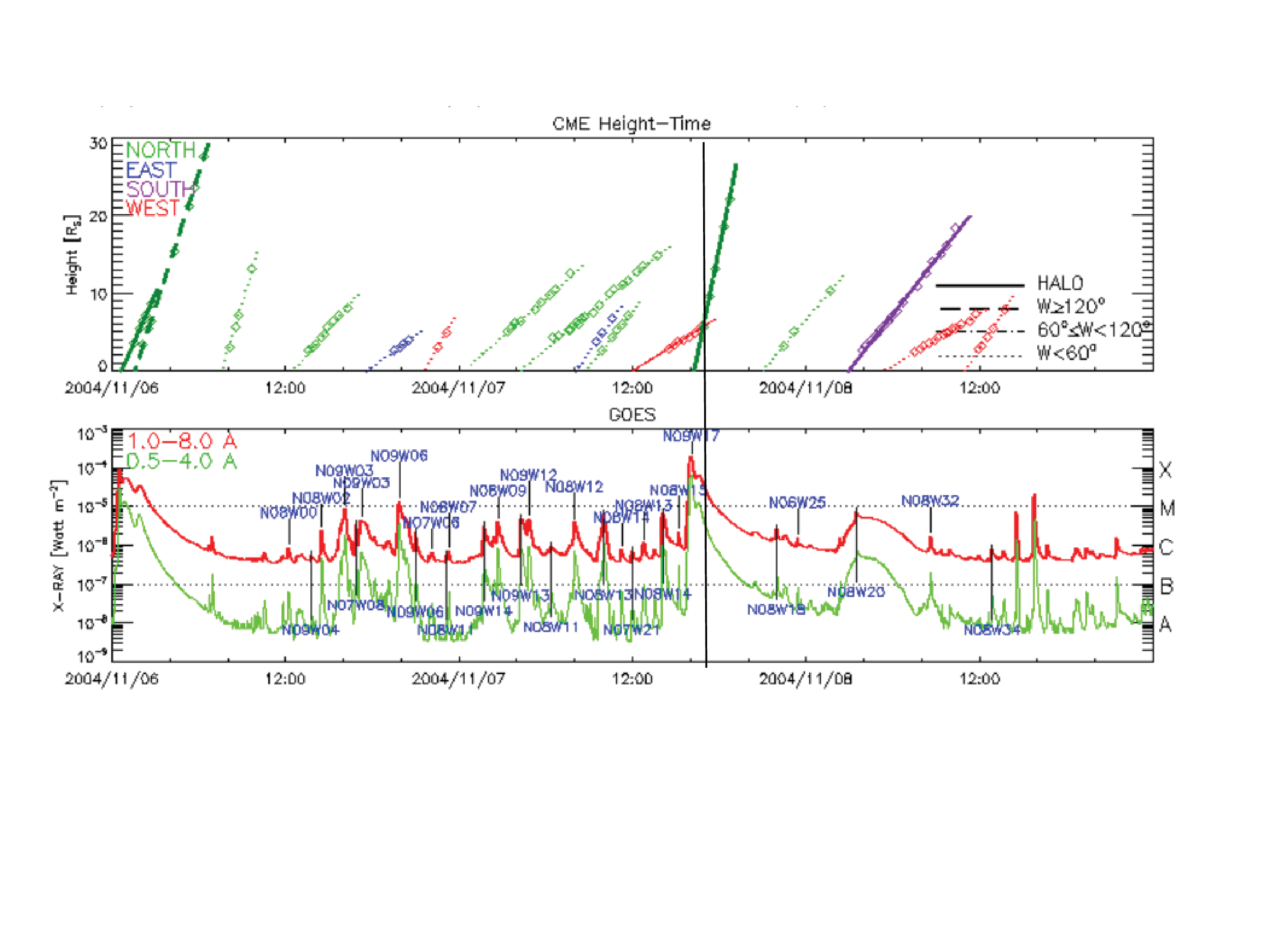} 
  \vspace{-20mm}
  \caption{ The CME emission log from November 6th 00 UT to November 8th 00 UT observed by the SOHO coronagraph.  
Two high speed CMEs are recognized in the plot in association with the two large solar flares at November 6th (at 00: 34UT) 
and November 7th (at 16 UT).  Before the emission of the CME at 16 UT, another CME was emitted around 12 UT of November 7th with rather slow speed.  
As a result, they will collide at 17 UT.  The collision time is indicated by the vertical line in the figure.  
The CME emission log is shown together with the GOES X-ray data as reference.  
The numbers in the GOES X-ray data indicate the flare position on the solar 
surface.  The data are available from SOHO home page.} 
\end{figure} 

\begin{figure}[ht] 
\vspace{-30mm}
  \includegraphics[width=120mm]{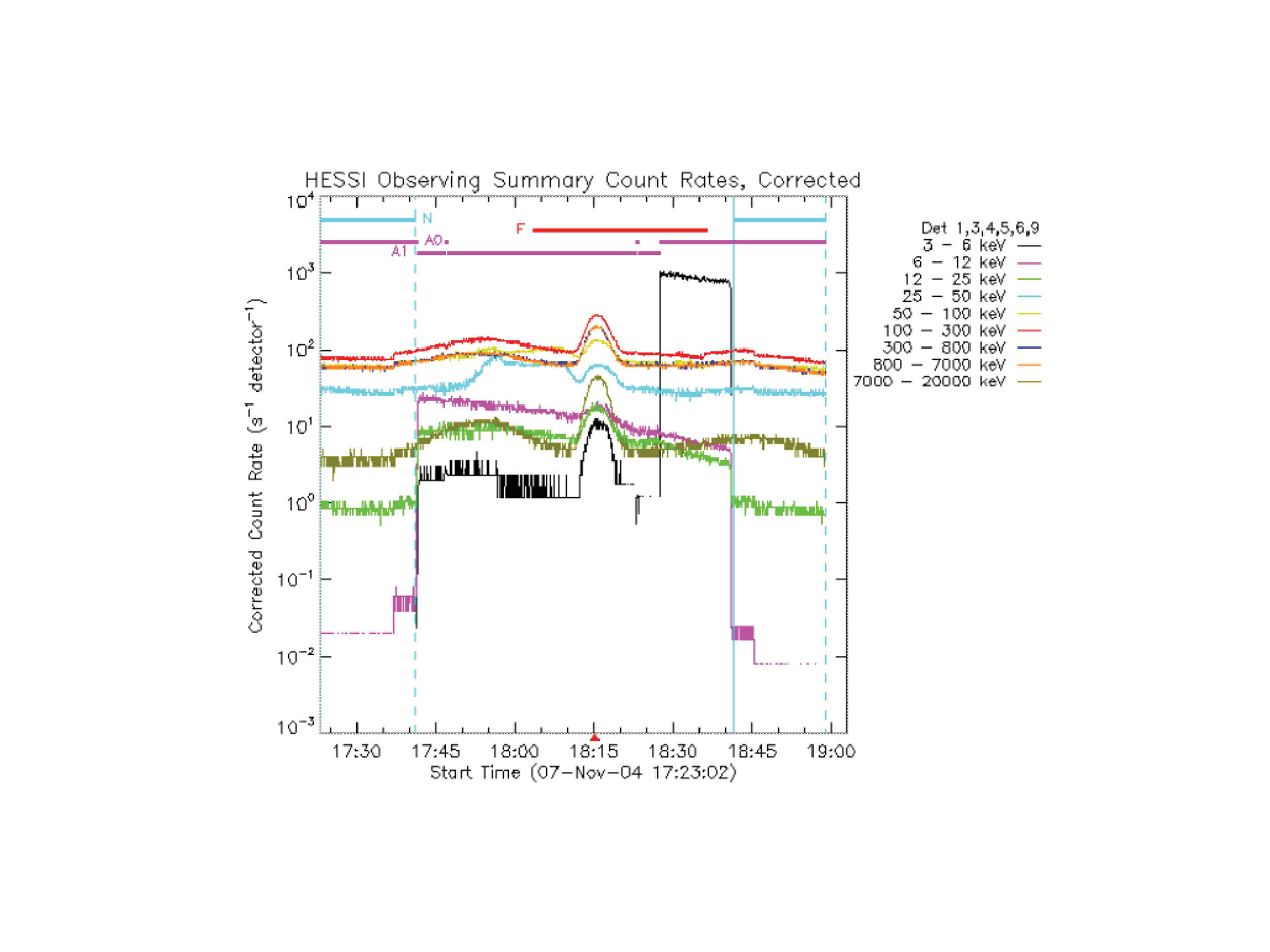} 
  \caption{The flux of X-rays observed by the RHESSI satellite.  At 18:15 UT a small bump can be recognized.} 
\end{figure}

\end{document}